# Design of Automated Dual B and 4G Jammer using MATLAB Simulink

**T. Chetan Sai, A. G. Dinesh Kumar, V. Charan and S. Ramya***

School of Electronics Engineering, IIIP, VIT University, Near Katpadi Road, Vellore - 632014, Tamil Nadu, India; ramya.sekar@vit.ac.in, chetansai.tutika@gmail.com, a.dineshkumar2014@vit.ac.in, vallapanenicharan@gmail.com

## Abstract

This paper presents a design of efficient smart jammer to jam the 4G signals, specifically band 3 and band 40 which can be predominantly used in India. The MATLAB Simulink tool was used for the analysis of the circuit design. Simulink, an advanced tool gives accurate results comparable to real time analysis. The DSP toolbox of the Simulink library has been largely used to construct and view the results of the model. The main objective of this paper is to receive the LTE signals, filter band 3 and band 40, add noise and increase the amplitude of the signal. The uniqueness of this design is the use of full wave rectifier in the circuit and the trigger enabled blocks. Full wave rectifier with added circuitry acts as a trigger to the jammer which contains the noise block and the gain block. Sine wave generators were used to replicate real time signals and additional signals were added as noise. The advantage of the design is when none of the bands are detected; the output will not be generated, thus saving power. The incorporation of the detector circuit and trigger circuit ensures that power is not wasted by the jamming circuit when there signal is not detected. These jammers can be used in examination halls, conference halls and in secure location where telecommunication signals are unwanted.

## 1. Introduction

Long Term Evolution (LTE), the fourth generation technology for mobile broadband suffers from loss of communication and data. The overhead linked with the practical applications are due to augmented signalling traffic. The unwanted users take advantage of this traffic and cause severe overload disturbing authentic users to access the network. To secure the communication technology, the jammers are invented for military and civilian applications. A mobile phone jammer prevents mobile phones receiving signals from the base stations. These devices are used in places wherever a phone call would be mostly troublesome[1]. There are other jamming devices like Intelligent Cellular Disablers, Intelligent Beacon Disablers, Direct Transmit and Receive Jammers and EMI Shield Passive Jamming for prohibiting mobile phones from ringing ins particular areas[2,3]. The mobile phone jammers limit the mobile phone usage by transmitting radio waves on same frequencies as mobile communication frequencies, producing interference and making the phones unusable[4,5]. There are numerous techniques to jam Radio Frequency (RF) signals like Spoofing, Shielding attacks, Denial of service etc[6-9]. MATLAB toolboxes such as DSP toolboxes have been used and also Graphical User Interface (GUI) can be used to display data[10]. Simulation has become a very important tool in analysing, optimizing and designing of circuits[11]. The toolboxes are mostly used for research purpose to handle and control the scheduling issues in real time systems. Using toolboxes it is possible to evaluate and modify the circuit design and hardware platforms to better understanding of the timing and performance[12].

** Author for correspondence*

In this paper, design and implementation of a new 4G dual band jammer for jamming mobile phone communication in 4G bands is presented. It blocks Band 3 (Frequency Division Duplexing (FDD) – LTE) and Band 40 (Time Division Duplexing (TDD) – LTE). There are various multiple access techniques like Time Division Multiple Access (TDMA) in which multiple users can send data in separate time slots in same frequency, Frequency Division Multiple Access (FDMA) where many users can send data at the same time in separate frequencies and the last one is Code Division Multiple Access (CDMA) where many users can send data at same time and in same frequency using different codes[13]. Frequency Division Multiplexing (FDM) channelizing is used to assign channels according to the patterns for least average energy cost[14]. The bands predominantly used in India, band 3 and band 40 are jammed effectively using this technique. The proposed technique is simple and cost effective. The designed jammer consists of detection circuit to detect the uplink frequency; once it is detected, the downlink signal is jammed. The trigger circuit is set up, to switch on the circuit only when signal is detected, thus reducing power conversion, the jamming circuit is triggered by the voltage from trigger circuit which adds noise and amplifies the signal before transmitting.

## 2. Design and Methodology

Primarily, any signal in the surroundings will be picked up by an antenna, which would then channel the signal into a suitable microcontroller containing the coding part of the design. The signals to be jammed are isolated by frequency filters. These signals, if present in the surroundings, trigger the modulator and act as an input to the modulating circuit. In these modulators, the input is modulated with a random noise signal and the power of resultant output signal is then amplified. The amplified signal is picked up by the electronic devices instead of the original signal and thus, the desired signal is jammed.

MATLAB Simulink was used for simulating the design. The proposed circuit design is divided into three parts namely Detection, Triggering and Jamming. The circuit comprises sine wave generators which represent the mobile signals present in real world, detection circuit (subsystem), trigger circuit (subsystem 1, subsystem 3), jammer circuit (subsystem 2, subsystem 6), time scopes and power GUI block. Signal passes through the filters in detection subsystem and are filtered according to the bandwidths. Separate filters are used to detect uplink and downlink frequencies of the band. The uplink frequencies are used as input to the trigger which outputs a constant DC voltage. The downlink frequencies are amplified and then added with noise and transmitted. The triggering circuit consists of a full wave rectifier, a switch and a 5V DC supply connected to the switch. The switch works as a comparator to compare input voltage with fixed threshold value. If the input is greater than threshold value, the output 5V supply is connected to it. The filtered signal after passing through the full wave rectifier are always more than 1V. On passing through the switch of threshold voltage 1V, outputs a 5V DC voltage if the amplitude of the rectified wave is greater than 1V or else the output is 0V. The voltage from the trigger circuit enables the jamming circuit. The 5V supply from the trigger circuit enables the subsystem (subsystem works only if it is triggered by positive voltage) which amplifies and adds noise to the signal and transmits it. If there is no trigger, the output will be 0V making the jammer circuit to be inactive.

If a particular frequency is detected by the filters, then the triggering circuit gets triggered and enables the jammer. Hence, power consumption is reduced. Figure 1 shows the overall circuit of the jammer. The detector circuit with filters tuned to the specific uplink and downlink frequencies of Band 3 (FDD-LTE) and Band 40 (TDD-LTE) are shown in Figure 2. For the 4G band 3, the uplink and downlink frequencies are 1710-1785 Hz and 1805-1880 Hz, respectively. For the 4G band 40, the uplink and downlink frequency is 2300-2400 Hz. Figure 3 shows the trigger circuit used to enable the jammer circuit. The Table 1 presents the stop band and pass band frequencies used by the filters used in detection circuit.

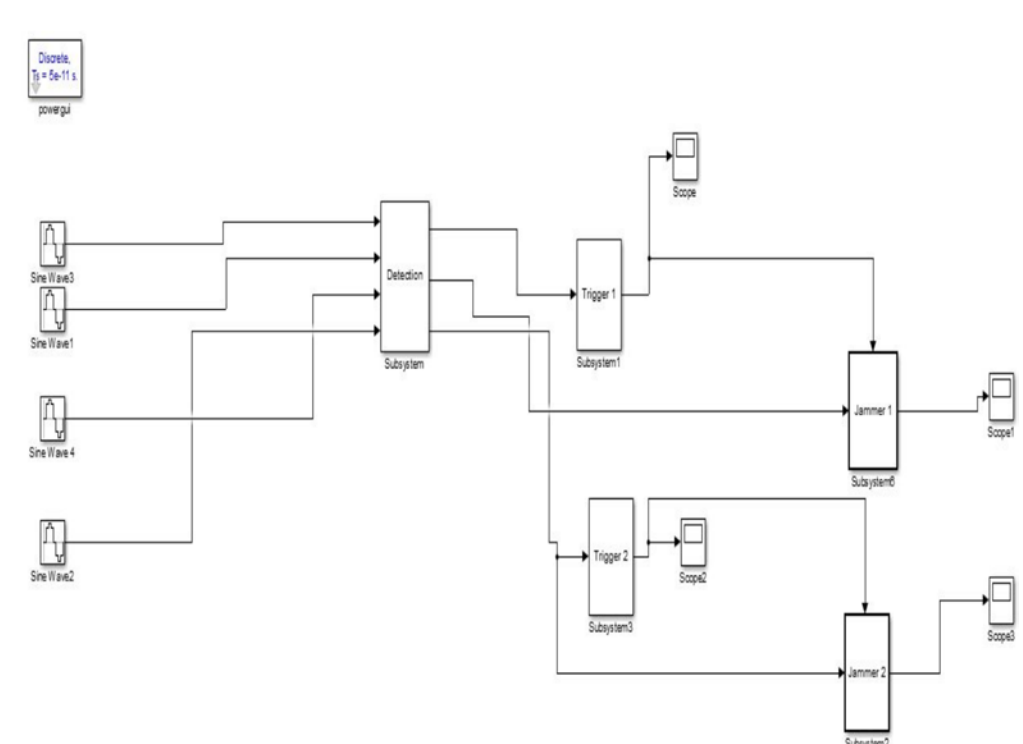

**Figure 1.** Circuit overview of the proposed Jammer..

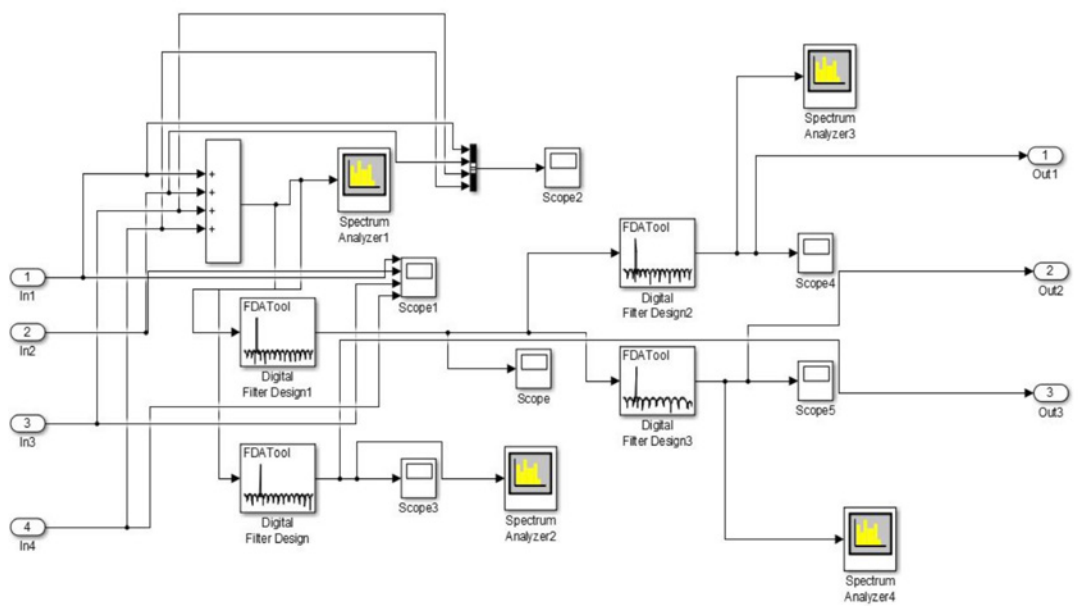

**Figure 2.** Detection circuit.

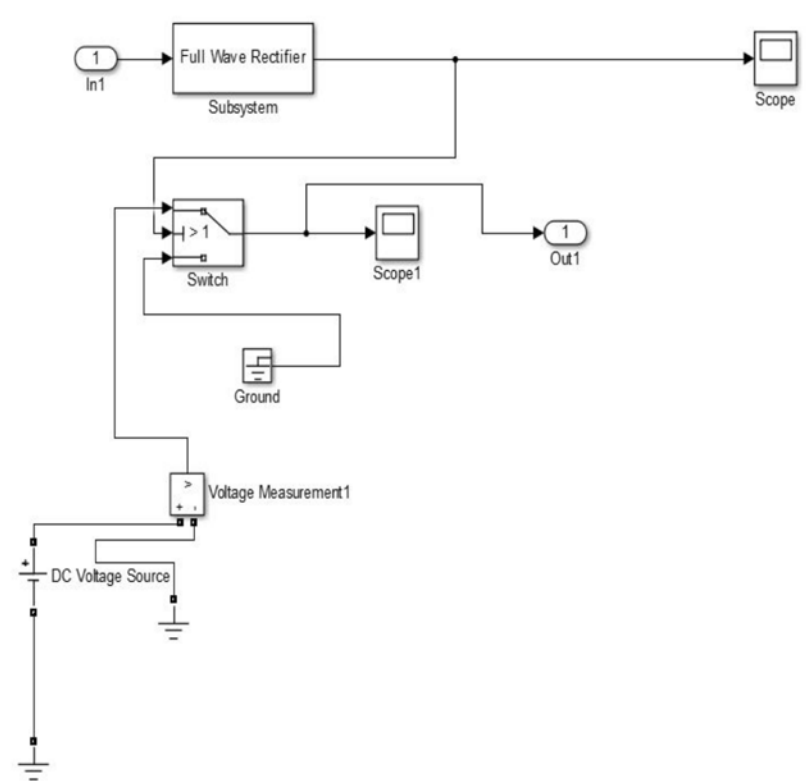

**Figure 3.** Trigger circuit.

**Table 1.** Cut off frequencies of the Filters

| Filter number | Band pass 1 (MHz) | Band pass 2 (MHz) | Centre frequency (MHz) |
|---|---|---|---|
| 1 | 1710 | 1880 | 1795 |
| 2 | 1710 | 1785 | 1747.5 |
| 3 | 1805 | 1880 | 1842.5 |
| 4 | 2305 | 2405 | 2355 |

## 3. Simulation Results and Discussion

In this section, the simulated results are presented and analysed. The Figure 4, Figure 5, Figure 6 and Figure 7 shows the input frequencies 1, 2, 3 and 4, respectively used for the simulation. Figure 4 (input 1) and Figure 5 (input 2) represents the signals with frequencies 1.2 GHz, 1.5 GHz, 1.6 GHz, 3GHz and 1.2GHz, 1.74 GHz, 1.85 GHz, 3GHz, respectively. Figure 6 (input 3) and Figure 7 (input 4) shows signals with frequencies 1.2 GHz, 1.3 GHz, 2.34 GHz, 3GHz and 1.74 GHz, 1.85 GHz, 2.34 GHz, 3GHz, respectively. Figure 8 is the output of jammer 1 and jammer 2 in voltage versus time graph by giving input 1 containing frequencies out of band 3 and band 40. The graph shows 0V as frequencies detected by the device are out of the dual bands 3 and 40. Figure 9 shows the output of jammer 1 when input 2 containing band 3 and random frequencies are used. It shows a distorted signal with higher voltage than the original signal. Jammer 1 displays output since it is associated with filter which detects band 3 of 4G spectrum. Figure 10 is the output of jammer 2, showing 0V output as the jammer 2 is associated with jamming of band 40 signals. Figure 11 is the output of trigger 1 of constant voltage of 5V which is used to trigger jammer 1.

The input 3 containing band 40 and random frequencies are applied to jammer and trigger and the results are analysed. Figure 12 represents the output of jammer 2 which shows a distorted signal with higher voltage than the original signal. The Jammer 2 shows output because it is associated with filter which detects band 40 of 4G spectrum. Figure 13 is the output of jammer 1 and it shows 0V output as jammer 1 is associated with jamming of band 3 signals. The output of trigger 2 showing a constant voltage of 5V which is used to trigger jammer 2 is presented in Figure 14. The input 4 which contains band 40, band 3 and random frequencies are applied and the results are shown in Figure 15. Figures 15 (a) and 15 (b) are the outputs of jammer 1 and jammer 2 showing distorted signal with higher voltage than the original signal. Jammer 1 and jammer 2 shows output because they are associated with filter detecting band 3 and band 40 of 4G spectrum, respectively. The simulation results clearly describe that the proposed 4G jammer blocks Band 3 and Band 40 efficiently.

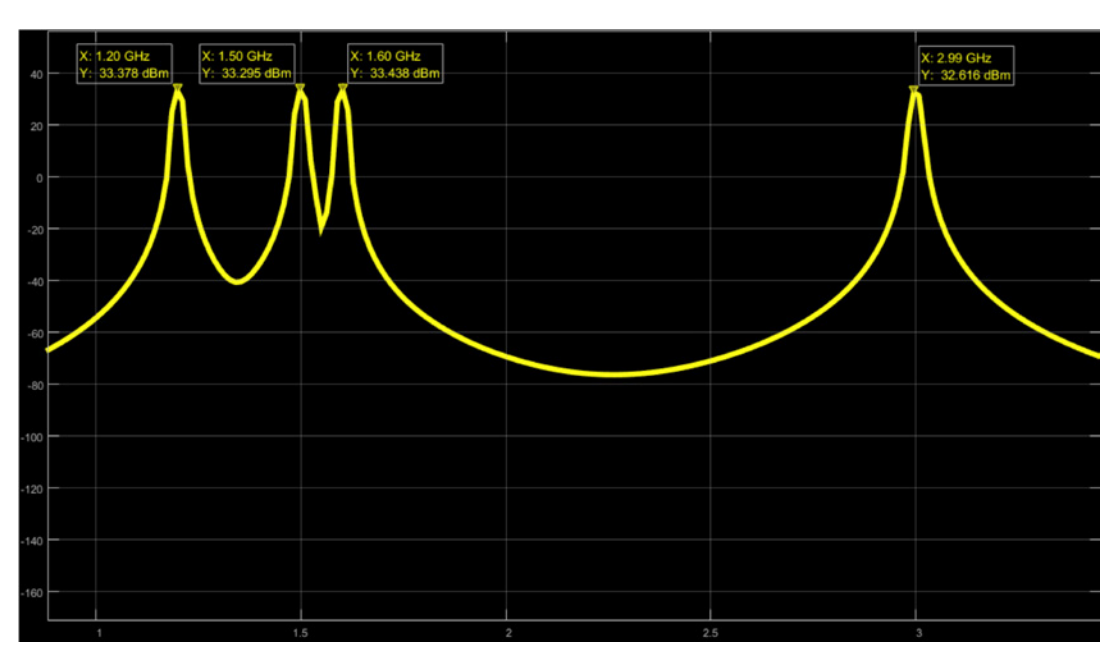

**Figure 4.** Input frequencies of values 1.2 GHz, 1.5 GHz, 1.6 GHz, 3 GHz. (Legend: x axis = frequency in GHz, y axis = power in decibels).

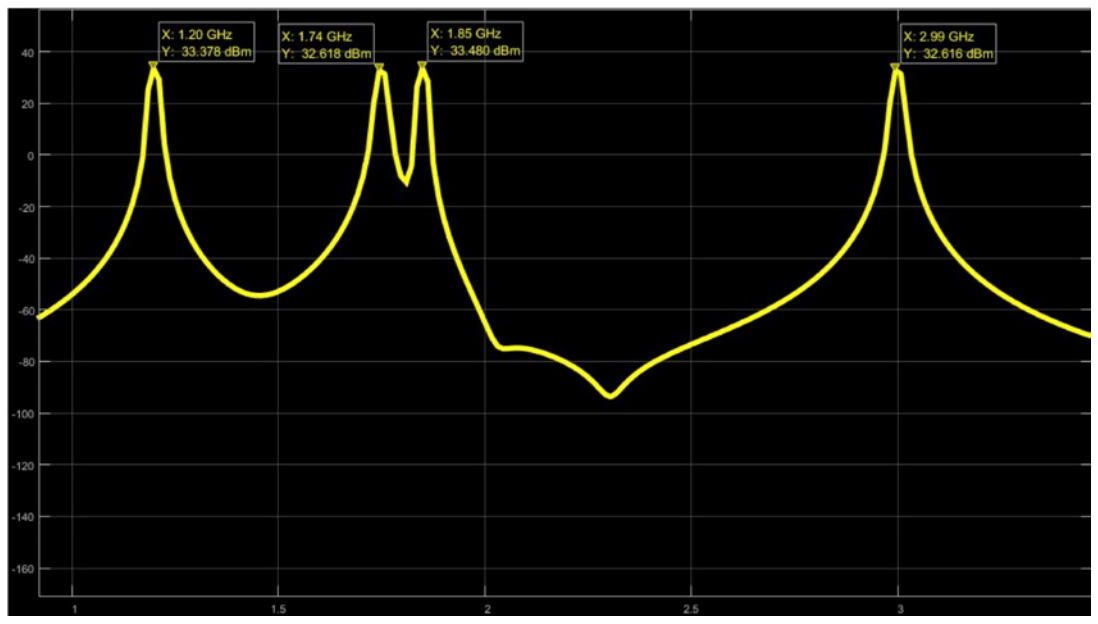

**Figure 5.** Input frequencies of values 1.2 GHz, 1.74 GHz, 1.85 GHz, 3 GHz. (Legend: x axis = frequency in GHz, y axis = power in decibels).

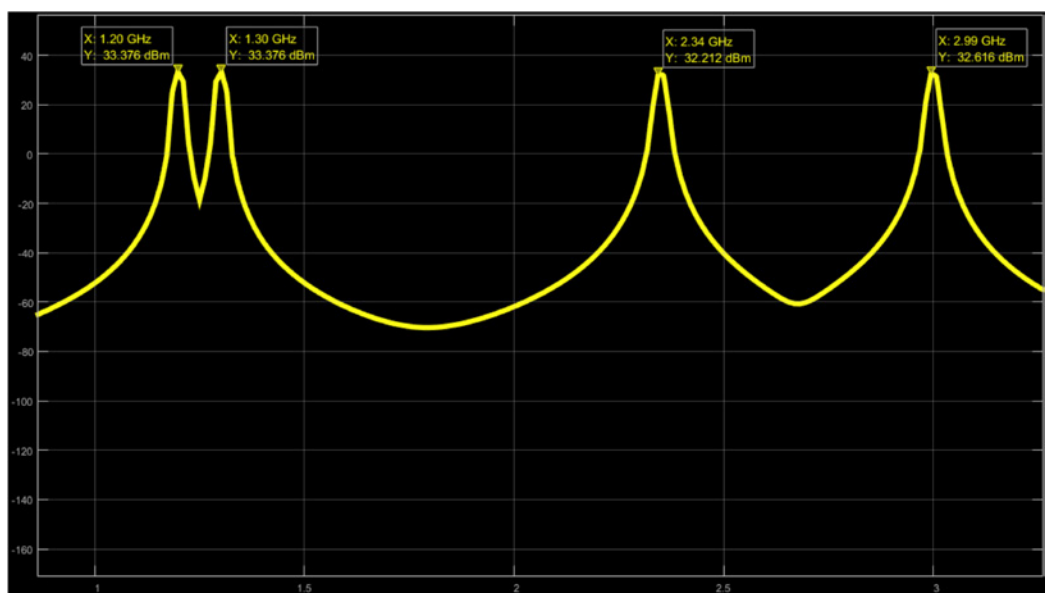

**Figure 6.** Input frequencies of values 1.2 GHz, 1.3 GHz, 2.34 GHz, 3 GHz (Legend: x axis = frequency in GHz, y axis = power in decibels).

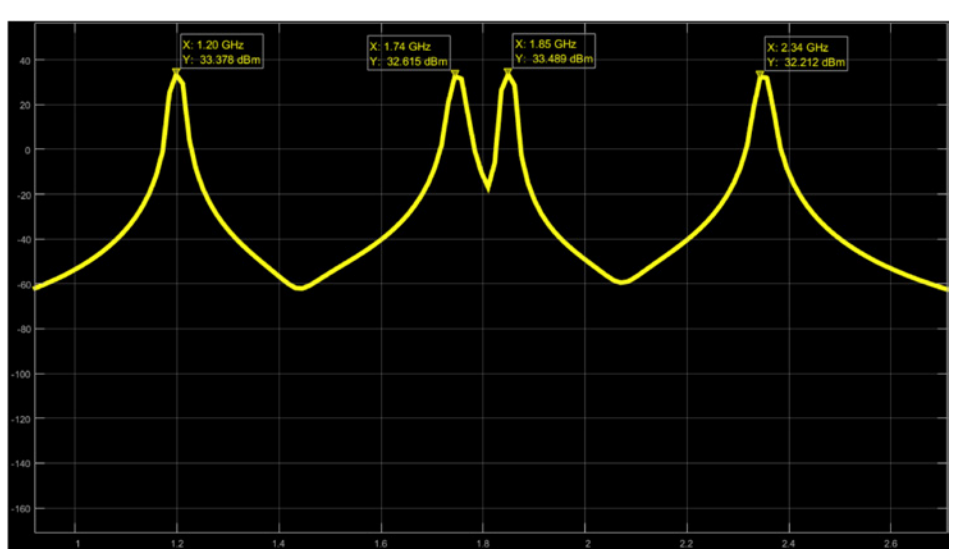

**Figure 7.** IInput frequencies of values 1.74 GHz, 1.85 GHz, 2.34 GHz, 3 GHz. (Legend: x axis = frequency in GHz, y axis = power in decibels).

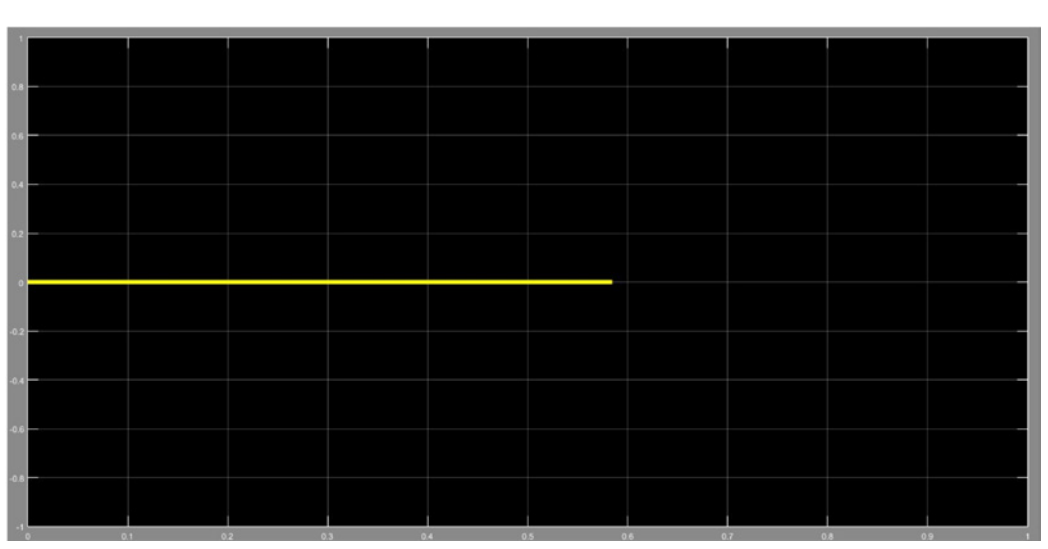

**Figure 8.** Output voltages of jammer 1 and jammer 2 for input 1 (Legend: x axis = time in 10-7 seconds, y axis = voltage in volts).

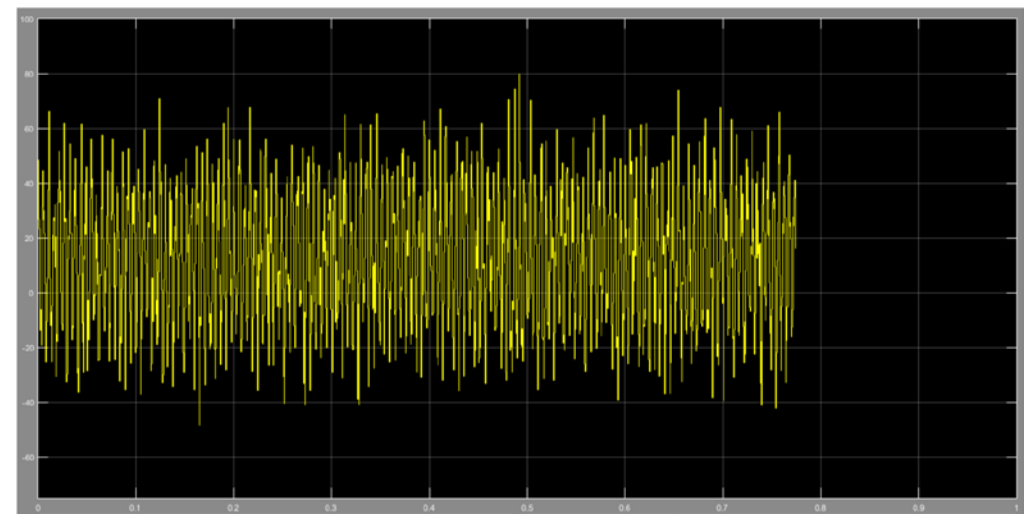

**Figure 9.** Output signal of trigger 1 for input 2. (Legend: x axis = time in 10-7 seconds, y axis = voltage in volts).

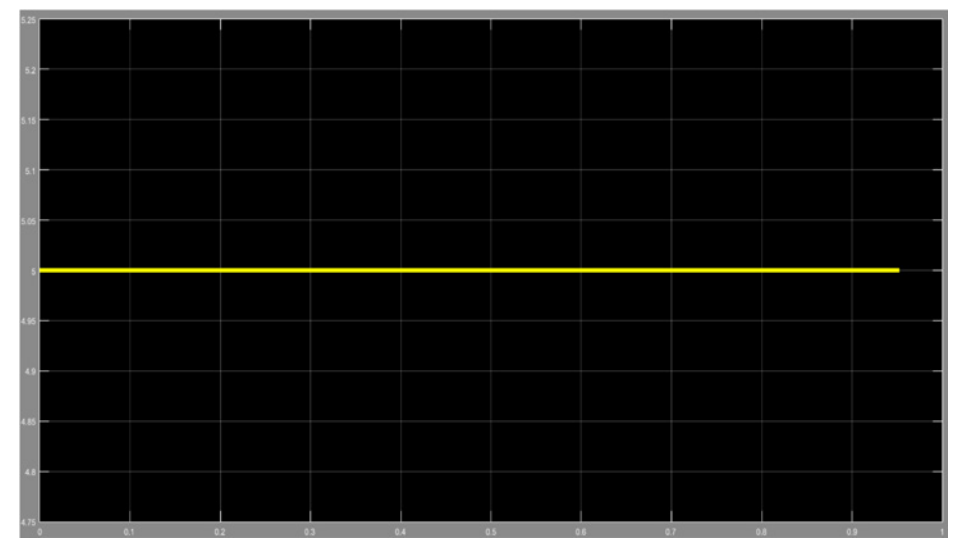

**Figure 10.** Output of jammer 1 for input 2. (Legend: x axis = time in 10-7 seconds, y axis = voltage in volts) in volts).

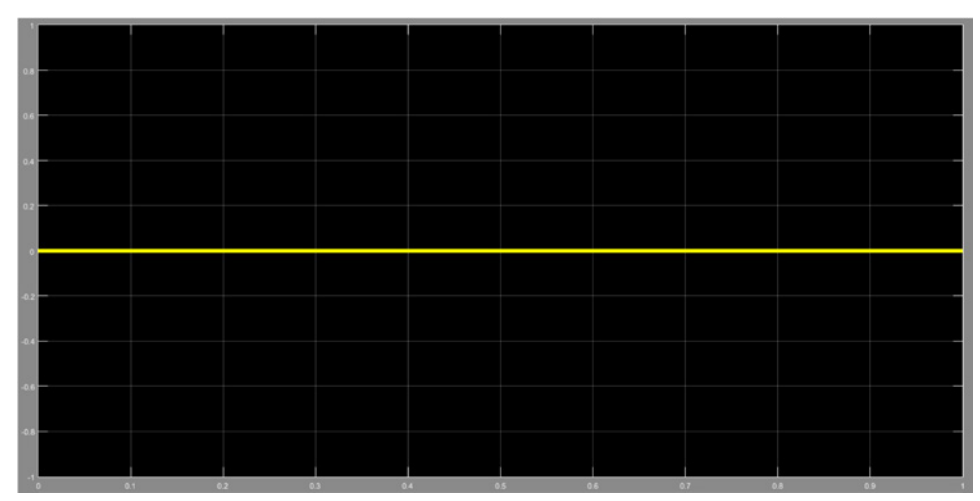

**Figure 11.** Output of jammer 2 for input 2. (Legend: x axis = time in 10-7 seconds, y axis = voltage in volts). in volts).

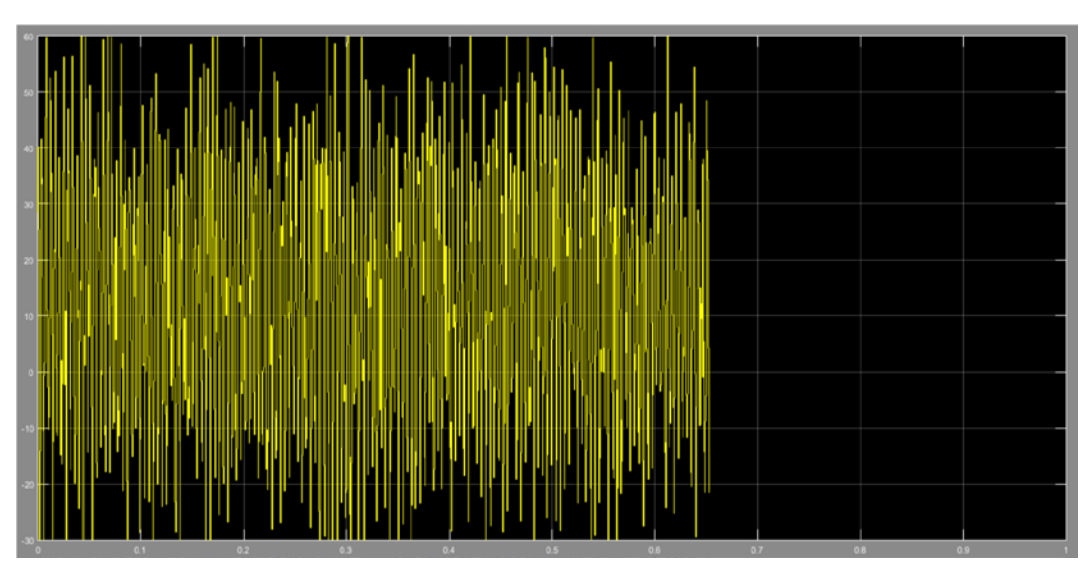

**Figure 12.** Output of trigger 2 for input 3. (Legend: x axis = time in 10-7 seconds, y axis = voltage in volts).

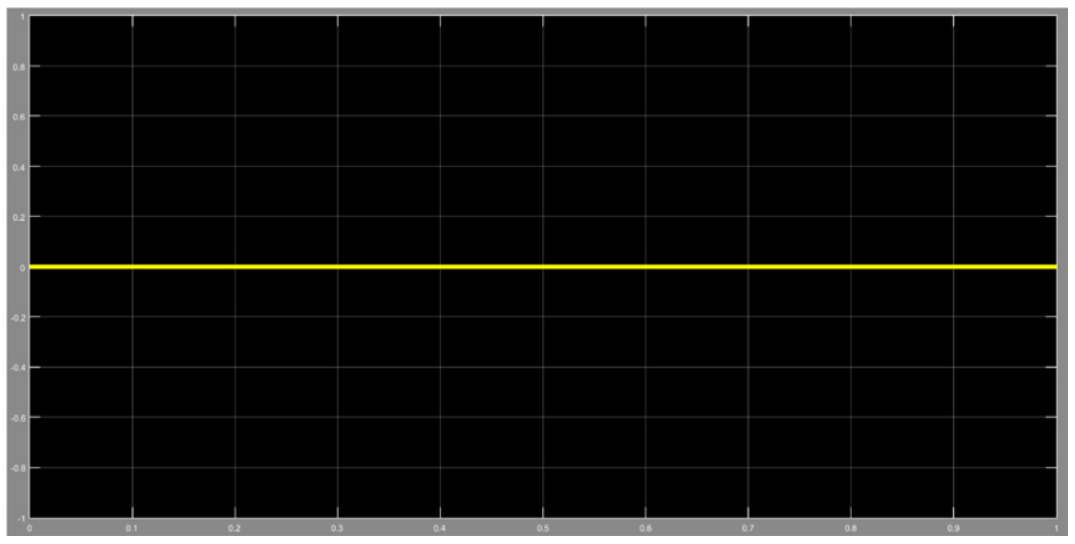

**Figure 13.** Output signal of jammer 2 for input 3. (Legend: x axis = time in 10-7 seconds, y axis = voltage in volts).

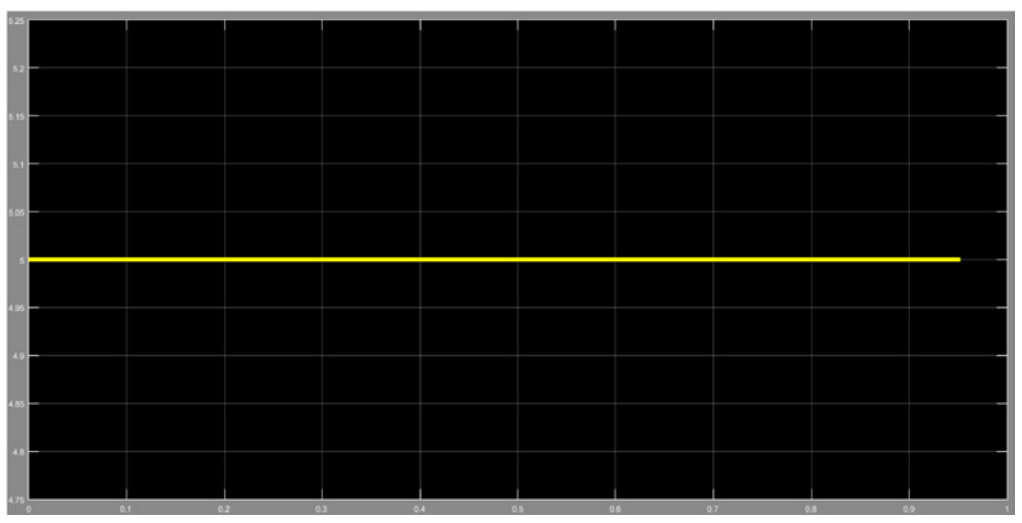

**Figure 14.** Output (0V) of jammer 1 for input 3. (Legend: x axis = time in 10-7 seconds, y axis = voltage in volts).

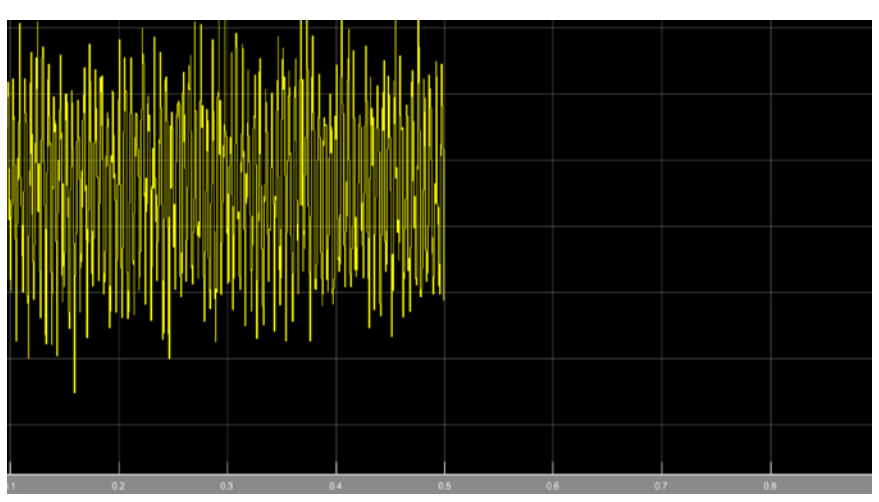

**Figure 15.(a)** Outputs of jammer 1 and jammer 2 for input 4. (Legend: x axis = time in 10-7 seconds, y axis = voltage in volts).

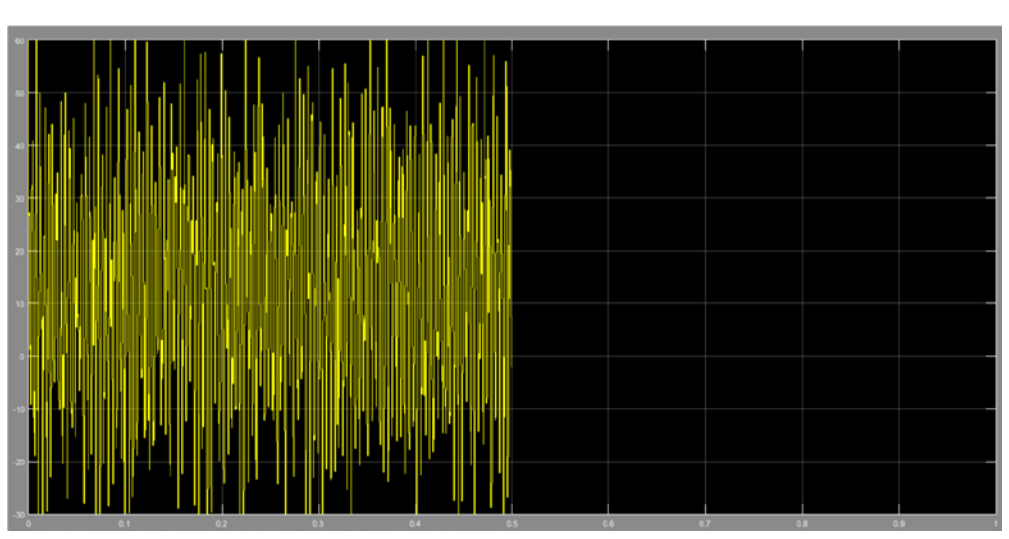

**Figure 15.(b)** Output of jammer 2 for input 4. (Legend: x axis = time in 10-7 seconds, y axis = voltage in volts).

## 4. Conclusion

In this paper, automated 4G jammer with new wave generators, detectors, filters, trigger and jamming circuit are designed and implemented using MATLAB SIMULINK based on the principle of Denial of service. The proposed jammer blocks LTE signals especially Band 3 and Band 40. The full wave rectifier is used for triggering the jammer circuit to jam the signals and is a promising method to significantly trigger the jammer circuit. The Gaussian and Rayleigh noises are added and finally jamming of LTE signals are achieved. The designed jammer is most promising in LTE mobile communication applications.

## 5. References

1. Vidyarani M, Sudhakar Y. Advanced Mobile Phone Signal Jammer for GSM, CDMA and 3G Networks with Prescheduled Time Duration using ARM 7. International Journal of Professional Engineering Studies. 2013 Dec; 1(2):207-10.
2. Gupta C, Garg C. Analysis of Jammer Circuit. International Journal of Engineering Research and General Science. 2014 Nov; 2(6):758-61.
3. Hussain A, Nazar N, Aqib A, Qamar S, Zia M, Mahmood H. Protocol-Aware Radio Frequency Jamming in Wi-Fi and Commercial Wireless. Journal of communications and networks. 2014 Aug; 16(4):397-406.
4. Krishnaiah RN, Brundavani P. FPGA Based Wireless Jamming Networks. International Journal of Modern Engineering Research. 2013 Aug; 3(4):2567-71.
5. Naresh P, Babu RP, Satyaswath K. Mobile Phone Signal Jammer for GSM, CDMA with Pre-scheduled Time Duration using ARM7. International Journal of Science, Engineering and Technology Research. 2013 Sept; 2(9):1781-84.
6. Nicholas W, Scott S. University of Nebraska: Study of cellular phone detection techniques. 2011.
7. Bhatia SK, Sharma K, Chaudhary K, Singh D. Signal Jammer and Its Applications. International Journal of Electrical and Electronics Research. 2015 Apr; 3(2):463-67.
8. Jisrawi A. GSM 900 Mobile Jammer. JUST. 2006; p. 1-28.
9. Mummoorthy A, Kumar SS. A Detailed Study on the Evolution of Recent Jammers in Wireless Sensor Networks. International Journal of Engineering Research and Development. 2012 Oct; 4(6):12-15.
10. Kumar R, Bansal K, Saini DK, Paul IPS. Development of Empirical Formulas and Computer Program with MATLAB GUI for Designing of Grounding System in Two Layer Soil Resistivity Model for High Voltage Air Insulated and Gas Insulated Substations. Indian Journal of Science and Technology. 2016 July; 9(28):1-7.
11. Bala I, Rana V. Performance Analysis of SAC Based Non-coherent Optical CDMA System for OOC with Variable Data Rates under Noisy Environment. Indian Journal of Science and Technology. 2009; 2(8):49-52.
12. Sivakumar P, Vinod B, Devi RS, Rajkumar ER. Real-Time Task Scheduling for Distributed Embedded System using MATLAB Toolboxes. Indian Journal of Science and Tech-

nology. 2015 July; 8(15):1-7.
13. Kumar CVR, Bagadi KP. Robust Neural Network based Multiuser Detector in MC-CDMA for Multiple Access Interference Mitigation. Indian Journal of Science and Technology. 2016 Aug; 9(30):1-7.
14. Munusamy K, Parvathi RMS, Chandramohan K. Least Power Adaptive Hierarchy Cluster Framework for Wireless Sensor Network using Frequency Division Multiplexing Channelization. Indian Journal of Science and Technology. 2016 Feb; 9(6):1-10.